\documentclass[twocolumn,showpacs,preprintnumbers,amsmath,amssymb]{revtex4}

\usepackage{graphicx}
\usepackage{dcolumn}
\usepackage{bm}
\begin{document}

\preprint{APS/123-QED}

\title{On the missing link between pressure drop, viscous dissipation and the turbulent energy spectrum}

\author{A. Badillo}
\email{a.badillo@imperial.ac.uk}
\affiliation{Department of Chemical Engineering,\\
Imperial College London, UK.}
\author{O. Matar}
\email{o.matar@imperial.ac.uk}
\affiliation{Department of Chemical Engineering,\\
Imperial College London, UK.}

\date{\today}

\newcommand{\Rey}{\mbox{\textit{Re}}}   
\newcommand{\vel}{\textbf{u}}
\newcommand{\stress}{\stackrel{\leftrightarrow}{\boldsymbol{\sigma}}}
\newcommand{\corr}{\stackrel{\leftrightarrow}{\textbf{R}}}
\newcommand{\fourier}{\hat{\textbf{R}}}
\newcommand{\deviatoric}{\stackrel{\leftrightarrow}{\boldsymbol{\tau}}}
\newcommand{\identity}{\stackrel{\leftrightarrow}{\textbf{I}}}
\newcommand{\Production}{\mathcal{P}}
\newcommand{\Dissip}{\mathcal{E}}

\begin{abstract}
We present convincing evidence of a direct connection between the pressure drop, viscous dissipation, and the turbulent energy spectrum. We use this finding to 
explain Nikuradse's experimental results of pressure drop for turbulent flows in rough pipes, in terms of a modified Kolmogorov length scale that varies with the surface roughness. Furthermore, we use Laufer's measurements of turbulent energy spectra in pipe flow to calculate---to a good approximation---the turbulent component of the pressure drop directly from an averaged turbulent energy spectrum. We also show that the incompressibility assumption, leads to the conclusion that viscous dissipation in fully-developed (laminar and turbulent) pipe flow, cannot increase the temperature of the fluid through viscous heating.
\end{abstract}

\pacs{}     
\maketitle


For many decades, turbulence has remained stubbornly one of the most challenging open problems in classical physics. Achieving complete understanding of turbulent flows is hindered by the their multi-scale nature, and the complex mechanisms underlying the distribution of energy to the large number of harmonics, which constitute the fluctuating velocity field. It is well accepted that vortex stretching is one of the main mechanisms responsible for transferring energy from low to high wavenumbers. This mechanism, however, only exists in viscous, three-dimensional turbulence, with no analogue in quantum  \cite{Barenghi}, and two-dimensional turbulence  \cite{Tran}. 

The pictorial energy cascade proposed by Richardson  \cite{Richardson}, and then recovered theoretically first by Kolmogorov  \cite{Kolmogorov1, Kolmogorov2}, and then, independently, by Onsager  \cite{Eyink}, does not reveal much about the energy transfer between consecutive harmonics; it only represents an `equilibrium' state of energy distribution arising from a balance between the fluxes of energy going up and down the cascade. Despite the fact that viscous dissipation is irrelevant for quantum turbulence, the scaling laws dominating its energy cascade are the same as those in viscous, three-dimensional turbulence  \cite{Maurer,Maltrud,Smith}. What, then, is the true role of viscous dissipation in the formation of the energy cascade, and what is its connection to macroscopic quantities such as the pressure drop? We address the second question in the present letter. 

\begin{figure}[htbp]
	\centering
		\includegraphics[width=8cm]{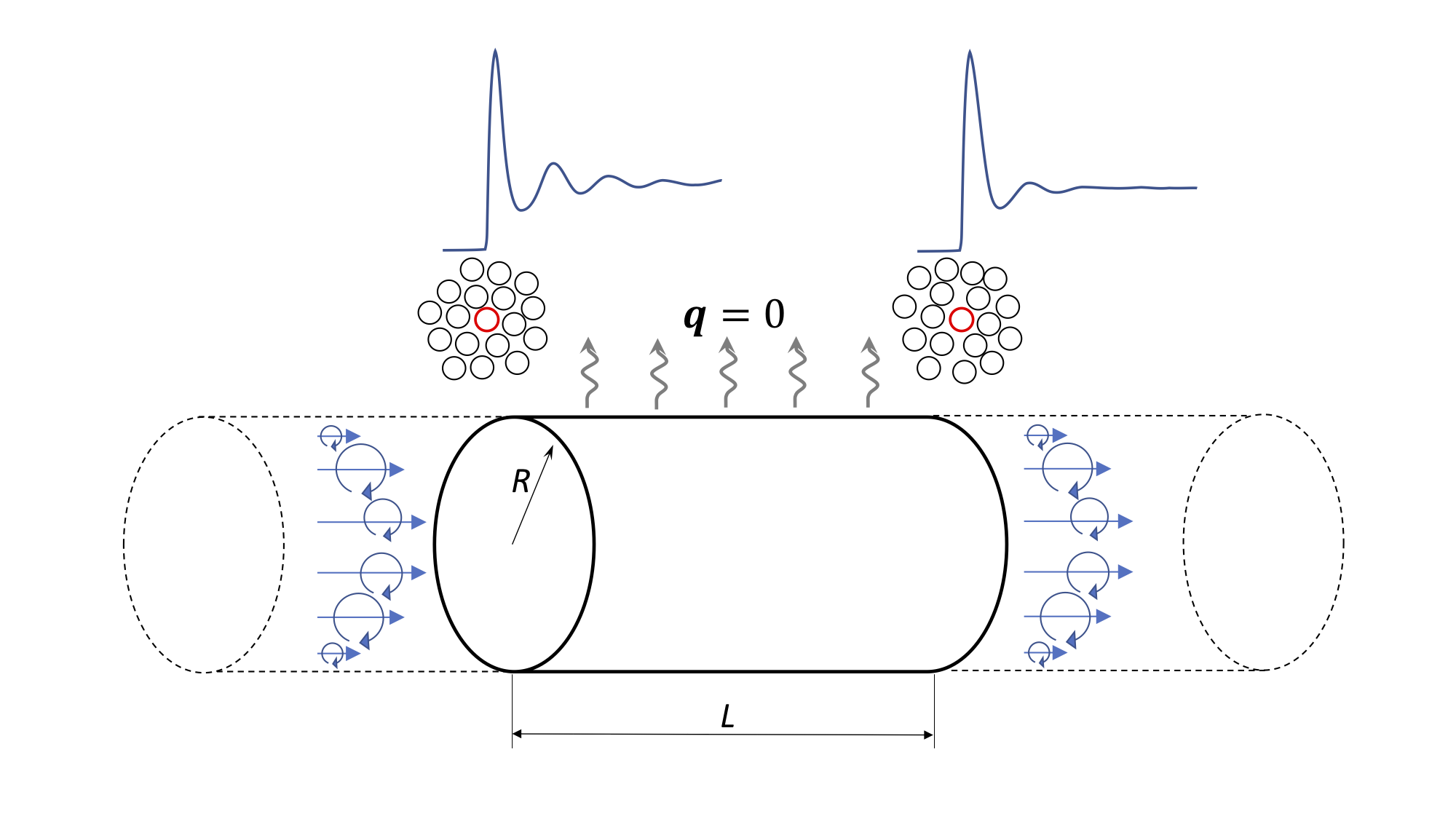}
	\caption{Schematic of fully developed pipe flow. The pipe is thermally insulated (adiabatic) and the fluctuations of any flow quantity are statistically identical at the inlet and outlet of the portion of the analyzed pipe. We also present how, hypothetically, the molecular radial distribution varies along the pipe (wavy lines in the upper part of the figure). The red circle surrounded by black ones represents, schematically, the local arrangement of atoms/molecules in the fluid. At inlet, the pressure is higher than that at the outlet, leading to a slightly higher local degree of order. This higher degree of short range order is observed as a larger amplitude of the second and subsequent peaks in the radial distribution.}
	\label{fig:schematicPipe}
\end{figure}

We start our discussion from the conservation of linear momentum \mbox{$\partial_t  \textbf{p} + \nabla\cdot \left(\textbf{u} \textbf{p} \right) = 
-\nabla\cdot \stress + \Sigma\textbf{f}_{ext}$}, with $\textbf{p}=\rho \textbf{u}$ the linear momentum, $\rho$ the density, $\textbf{u}$ the instantaneous velocity field, $\stress$  the stress tensor, and  $\Sigma\textbf{f}_{ext}$ the summation of all external forces. We further split the stress tensor as $\stress = P\identity - \deviatoric$, with $P$ the pressure and $\deviatoric = \mu \left( \nabla \textbf{u} + \nabla \textbf{u}^T\right)$ the viscous stress tensor expressed in terms of the dynamic viscosity $\mu$ and the strain rate. In our analysis, we consider a portion of a pipe, where the flow is adiabatic, incompressible and fully developed (Fig. \ref{fig:schematicPipe}). Hence, all the quantities are statistically invariant along the stream-wise direction, except for the fluid pressure.

By applying a time average to the conservation of momentum, we arrive to the following expression for the pressure drop
\begin{equation}
\Delta P = -\frac{2L}{R}\rho u_{\tau}^2
\label{eqn:pressure_drop_1}
\end{equation}
with $R$ the pipe radius, $L$ the pipe length, $u_{\tau}=\sqrt{\tau_w/ \rho}$ the friction velocity, \mbox{$\tau_w=\left(\mu \partial U / \partial r\right)_w$} the wall shear stress, and $U$ the stream-wise component of the time-averaged velocity field. 

An equivalent expression can be obtained from the conservation of kinetic energy $e_k = \textbf{u}^2/2$ (per unit mass) \mbox{$\partial_t \left(\rho e_{k}\right) 
+\nabla\cdot \left( \textbf{u}\rho e_{k}  + 
\stress\cdot\textbf{u} \right) = 
\stress\colon\nabla\textbf{u}
+\Sigma\textbf{f}_{ext}\cdot\textbf{u}$}. By applying a time average to the conservation law for kinetic energy and invoking incompressible and fully developed pipe flow conditions, the pressure drop reads
\begin{equation}
\Delta P = -\frac{1}{V A}\int_{\Omega}\left\langle \deviatoric\colon\nabla\textbf{u} \right\rangle d\Omega,
\label{eqn:pressure_drop_2}
\end{equation}
with $\left\langle \cdot \right \rangle$ a time average, $VA=\int_A \textbf{u}\cdot d\textbf{A}$ the volumetric flow rate, $A$ the pipe cross section, and $V$ the average cross section velocity (see supplemental material for a detailed derivation of conservation laws and the two expressions for the pressure drop). Equation (\ref{eqn:pressure_drop_2}) was also presented by Pope \cite{Pope}. Since the conservation law for kinetic energy is not independent from that of momentum, Eqs. (\ref{eqn:pressure_drop_1}) and (\ref{eqn:pressure_drop_2}) are completely equivalent. Thus, we can express the wall shear stress in terms of the total viscous dissipation.
\begin{equation}
\tau_w = -\frac{1}{2 \pi R L V }\int_{\Omega}\left\langle \deviatoric\colon\nabla\textbf{u} \right\rangle d\Omega,
\label{eqn:wallShear}
\end{equation}
We need to emphasize that Eq. (\ref{eqn:wallShear}) is only valid for fully developed (laminar or turbulent) and incompressible pipe flow. For laminar flow, it is possible to confirm analytically the validity of Eqs. (\ref{eqn:pressure_drop_2}) and (\ref{eqn:wallShear}). The first interesting conclusion from Eq. (\ref{eqn:pressure_drop_2}), is that viscous dissipation cannot increase the temperature of a fluid at the incompressibility limit. The conservation of enthalpy per unit mass, $h$, implies that \mbox{$\rho D_{t} h +\nabla\cdot\textbf{q}=  D_{t} P+\deviatoric\colon\nabla\textbf{u}$} (see supplemental material), with $D_t = \partial_t + \textbf{u}\cdot\nabla$ the material derivative and $\textbf{q}$ the heat flux. If the pipe is adiabatic and the flow incompressible, the only possibility for the enthalpy to change, is through viscous heating (dissipation term). After applying a time average to the conservation law of enthalpy, the integration of the R.H.S. over the pipe volume, vanishes by virtue of Eq. (\ref{eqn:pressure_drop_2}) and, thus, $\int_\Omega \left \langle \rho D_t h \right\rangle d\Omega = 0$. Since there is no energy exchange with the environment (adiabatic flow), this requires either a constant fluid temperature or that the local viscous dissipation increase and decrease the local fluid temperature in a way that the net change of total enthalpy is null. Because the second option would be in violation with the first law of thermodynamics, the fluid temperature must be constant at the incompressibility limit. 

For molecular fluids like water, the internal energy can suffer important variations by changing the molecular arrangement, keeping an almost constant temperature and density. A small change in the enthalpy per unit volume, $\tilde{h}$, is given by $d\tilde{h} = d e_u + dP$, with $e_u$ being the internal energy per unit volume. If the enthalpy of the fluid is constant, then $d e_u/dP = -1$. Molecular Dynamics (MD) simulations of water at constant temperature  \cite{Mahoney}, indicate that the local range order (a measure of how molecules are arranged in the liquid) undergoes important changes as pressure increases. Using the MD data from Mahoney and Jorgensen  \cite{Mahoney}, we can estimate the derivative $d e_u/dP$. Applying a linear fit to their MD data yields a value of $d e_u/dP \sim -0.826$ (see supplemental material). Since the fluid considered in the MD study is slightly compressible  \cite{Mahoney}, we would expect a value slightly different from $-1$. These molecular studies, show the strong degree of correlation between internal energy and the molecular arrangement of the liquid. This would imply that for strictly incompressible fluids, the energy of the flow (pressure + kinetic) is dissipated only into the potential component of the internal energy following a reversible path, and not, as commonly believed, into the micro-kinetic component of the internal energy (temperature). For a liquid flowing in a pipe, the higher pressure at the inlet will cause a slightly higher degree of molecular order compared to that at the outlet (see Fig. \ref{fig:schematicPipe}). 

To reveal the connection between Eq. (\ref{eqn:pressure_drop_2}) and the turbulent the energy spectrum, we follow the work of Gioia and Chakraborty\cite{Gioia1}. They investigated the relationship between the friction factor, $f = 8\tau_w/\rho V^2$ and the phenomenological turbulent energy spectrum. Considering the Darcy-Weisbach equation, we can express the pressure drop as
\begin{equation}
	\Delta P = -f \frac{L}{D_H} \frac{\rho V^2}{2},
	\label{eqn:darcy-weisbach}
\end{equation}
where $D_H$ the hydraulic pipe diameter. The negative sign is to enforce a decrease in the pressure along the pipe. By replacing Eq. (\ref{eqn:pressure_drop_2}) into (\ref{eqn:darcy-weisbach}) and introducing the Reynolds decomposition $\textbf{u}=\textbf{U} + \textbf{u}'$, with $\textbf{U} = \left\langle \textbf{u}\right \rangle$, we can express the friction factor in terms of two components of the viscous dissipation,

\begin{equation}
f = \frac{64}{\Rey}\frac{1}{8\pi L V^2}\left(
\int_{\Omega}\frac{\overline{\epsilon}_0}{\nu}d\Omega
+\int_{\Omega}\frac{\epsilon'_0}{\nu}d\Omega
\right) 
\label{eqn:friction}
\end{equation}
where $\nu$ is the kinematic viscosity, and \mbox{$\rho^{-1}\left\langle \deviatoric\colon\nabla\textbf{u}\right\rangle = \overline{\epsilon_0} + \epsilon_0'$} is defined as the total time-averaged dissipation, with $\overline{\epsilon_0}/\nu = \nabla\textbf{U}\colon\nabla\textbf{U}+\nabla\textbf{U}^T\colon\nabla\textbf{U}$ and $\epsilon_0'/\nu = \left\langle \nabla\textbf{u}'\colon\nabla\textbf{u}' + \nabla\textbf{u}'^T\colon\nabla\textbf{u}'\right\rangle$ the mean and turbulent components of the viscous dissipation respectively. The Reynolds number, $\Rey = V D_H/\nu$, is defined in terms of the hydraulic pipe diameter, $D_H$ . 

The main difficulty faced by Gioia and Chakraborty \cite{Gioia1}, was the determination of the wall shear stress, and how to relate it to the turbulent energy spectrum and the surface roughness. They postulated that the wall shear stress can be expressed in terms of a momentum transfer between the bulk flow and the flow near the wall, characterized by a velocity scale $u_s$. Expressing the wall stress as $\tau_w=\rho V u_s$, the problem reduces to finding the velocity scale, which might also vary with the amplitude of the surface roughness. Furthermore, to determine the velocity scale they assumed the validity of the phenomenological energy spectrum in the proximity of the wall, which is something arguable. It is well known, however, that to derive the phenomenological spectrum, the turbulent structures must be rotationally- and translationally-invariant; that is, the flow is isotropic and homogenous, which is not the case near the wall. Aware of this limitation, Gioia and Chakraborty\cite{Gioia1} argue that even for anisotropic and inhomogeneous flows, the phenomenological theory still represents a good approximation. Thus, by using directly the energy spectrum $E\left(q\right)$ (including corrections for the energetic and dissipative ranges), they obtain the velocity scale as $u_s^2 = \int_{1/s}^{\infty}{ E\left(q\right)}dq$. Here, the integration is carried out for wave numbers higher than $1/s = 1/\left(r+a\eta\right)$, where $a$ is a dimensionless constant, $r$ the size of the surface roughness, and $\eta$ the Kolmogorov length scale. Hence, they are effectively filtering out all eddies larger than $s$ in the calculation of the velocity scale. It is remarkable that for $\Rey \rightarrow \infty$, the friction factor in the work of Gioia and Chakraborty\cite{Gioia1} scales as $f \sim (r/R)^{1/3}$ despite the absence of a correction for the dissipative regime, $C_d\left(q\right)$, in the turbulent energy spectrum: without this, the total dissipation diverges when using the Kolmogorov spectrum. The authors recovered the proper scaling for the friction factor, observed in Nikuradse's experimental results, only by selecting an appropriate cutoff wave length ($1/s$) in the integration of the energy spectrum. 

To avoid an arbitrary selection of the cutoff wave length, we follow a different route. In our analysis, we start by investigating the direct connection between the friction factor, defined in Eq. (\ref{eqn:friction}) with the turbulent energy spectrum. Considering an spherically and volumetric averaged spectrum $\hat{R}_0$ (see details in the supplemental material), along with Taylor's identities for isotropic turbulence\cite{Taylor}, we can approximate the total viscous dissipation by \mbox{$\epsilon'_0\approx 15\nu\int_0^\infty q^2 \hat{R}_0 dq $}. It is noteworthy to mention that in our analysis, $\hat{R}_0$, does not correspond to the spectrum of any particular point inside the pipe, but rather an averaged quantity.
We now analyze the asymptotic behavior of Eq. (\ref{eqn:friction}). For $\Rey\rightarrow 0$, $\int_{\Omega}\overline{\epsilon}_0/\nu d\Omega\rightarrow 8\pi L V^2 $ and $\epsilon'_0 \rightarrow 0$. Thus, the appropriate scaling for laminar flow $f\sim\Rey^{-1} $ is recovered directly. Conversely, for $\Rey\rightarrow \infty$, $\int_{\Omega}\overline{\epsilon}_0/\nu d\Omega\rightarrow 0 $ and $\int_{\Omega}\epsilon'_0/\nu d\Omega\sim L V^2 \Rey$, which is a result obtained from imposing a constant friction factor for high Reynolds numbers. The question that remains to be answered now is whether the two scalings, observed in rough pipe flows, $f\sim\Rey^{-1/4}$ and $f\sim \left(r/R\right)^{1/3}$ as $\Rey \rightarrow \infty$, can be explained solely in terms of the fluctuating component of the viscous dissipation. We, therefore, center our attention on the second integral in Eq. (\ref{eqn:friction}). Since we are interested in the dissipative range of the spectrum, we consider only the correction for this regime $C_d\left(q\right)$. 
Because $\hat{R}_0$ is isotropic and homogeneous, we will also make use of a modified Kolmogorov spectrum to determine this integral. Hence, we choose $\hat{R}_0 = A_0 \tilde{\epsilon}^{2/3}q^{-5/3}C_d\left(q\right)$, with $A_0$ a dimensionless constant and $\tilde{\epsilon}$ a dissipation rate scale (see supplemental material for its derivation) that must not be confused with $\epsilon'_0$. Thus, $\int_{\Omega}\epsilon'_0/\nu d\Omega \approx 15\Omega A_0 \tilde{\epsilon}^{2/3}\int_0^\infty q^{1/3} C_d\left(q\right) dq$. It is now clear that without the correction for the dissipative regime, the total dissipation diverges when using a Kolmogorov spectrum, which would translate into an infinite friction factor. 
\begin{figure}[htbp]
	\centering
		\includegraphics[width=8.5cm]{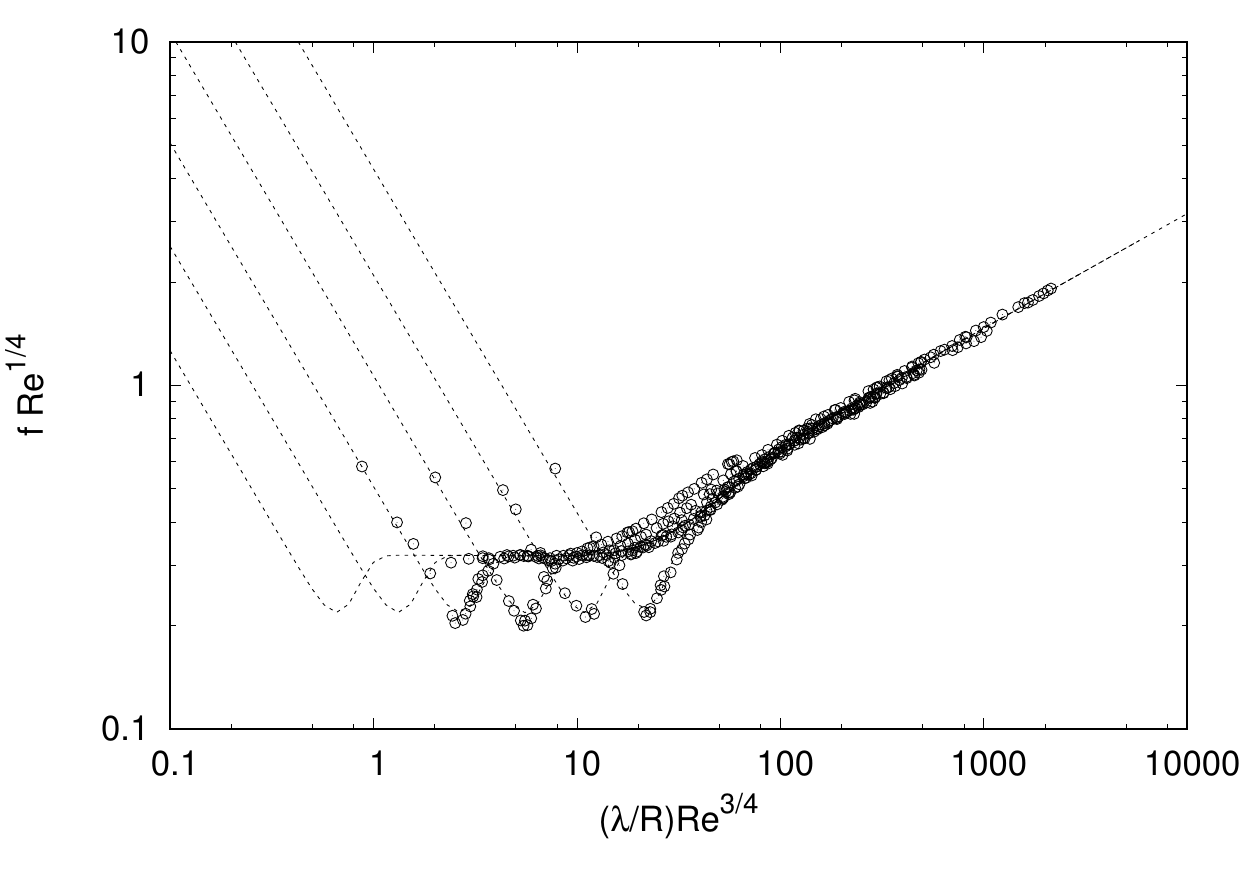}
\caption{Reproduction of Nikuradse's data over the entire range of Reynolds number.}
\label{fig:Nikuradse}
\end{figure}
To recover the proper scaling for the turbulent friction factor, the problem reduces to choosing an appropriate correction $C_d(q)$. For instance, we can recover the correct scaling by selecting $C_d(q)=\exp\left(-\beta \psi\left[R/\eta,\Rey, r/R\right] \eta q \right)$, with $\psi\left[R/\eta,\Rey, r/R\right]$ a dimensionless function, $\eta = \nu^{3/4}\tilde{\epsilon}^{-1/4}$ the Kolmogorov's length scale, $\beta=\left(30A_0 I_0 C_{\epsilon}^{11/12}\right)^{3/4}$ a dimensionless constant, and $I_0$ and $C_{\epsilon}$ constants determined in the supplemental material. By choosing \mbox{$\psi=\left(2R/ \eta\right)^{1/4}\left(A_2\left(1-\phi_2\right)+A_3\phi_2\left(\Rey^{3/4}\left(r/R\right)\right)^{1/3}\right)^{-3/4}$} (see supplemental material) we recover the scaling $f_T\sim\Rey^{-1/4}$ and $f_T\sim \left(r/R\right)^{1/3}$. The blending function $\phi_2=1-\left[1+\left(z/z_0\right)^2\right]^{-1}$, with $z=\Rey^{3/4}\left(r/R\right)$, bridges the Blasius and Strickler regimes. The constants $z_0=35.5$, $A_2 = 0.32 $, $A_3=0.147$ were determined from a fit to Nikuradse's data. Goldenfeld\cite{Nigel} postulated that the turbulent friction factor can be expressed in general as $f_T=\Rey^{-1/4}G\left(\Rey^{3/4}\left(r/R\right)\right)$ for the Blasius and Strickler regimes, with $G$ an unknown function of the Reynolds number and the relative surface roughness $r/R$. By plotting Nikuradse's data in a plot of $f_T\Rey^{1/4}$ vs $\left(r/R\right)\Rey^{3/4}$, Goldenfeld was able to collapse almost perfectly the experimental points onto a single curve. By utilizing a second function $\phi_1 = 0.5\left(1 + \tanh\left[\left(\Rey-\Rey_T\right)/\alpha_1\right] \right)$ to bridge the laminar and turbulent regimes, with $\Rey_T = 2944.27$ and $\alpha_1=1089.43$ (also determined from a fit), we can reproduce Goldenfeld's findings over the entire range of $\Rey$ including the laminar regime. Thus, by expressing the friction factor as $f = f_L \left(1-\phi_1\right) + f_T \phi_1$, with $f_L$ the friction factor in the laminar regime, we can closely match Nikurade's results (see Fig. \ref{fig:Nikuradse}). Nikuradse's data contain values from the laminar, transitional and turbulent regimes. To avoid an arbitrary selection of the data in the fully turbulent regime, we included the full data set in the fit. The value determined for $\Rey_T$, coincides well with the Reynolds number for transitional turbulent flows.

We have shown that a modified Kolmogorov spectrum, with a variable dissipative length scale (which depends on the surface roughness), can reproduce well Nikuradse's data. However, we must still provide evidence that an isotropic and homogeneous spectrum can be used to determine the turbulent component of the pressure drop. Finding a complete data set (experimental or numerical) to obtain an averaged spectrum has proven difficult, because we need spectra of velocity fluctuations measured along different directions at several locations inside a pipe. So far, we have found only one public data base that contains the spectrum for the three components of the fluctuating velocity field at various distances from the wall. To calculate an averaged spectrum for fully developed pipe flow, we use the measurements of Laufer\cite{Laufer}. Newer data such as the one produced in the Superpipe at Princeton University \cite{Smits}, only presents the spectrum for the streamwise component of the fluctuating velocity field. Looking at the individual spectra presented by Laufer (see supplemental material), we can see that at low wave numbers, the streamwise fluctuations, $u'u'$, contain a much higher energy level than the other two spanwise velocity components $v'v'$ and $w'w'$ (at a given distance from the wall). This means that most of the turbulence kinetic energy is contained in the streamwise velocity component $u'u'$. However, the situation is quite different at high wave numbers, where it is the spanwise fluctuations that contain most of the energy. Since viscous dissipation is relevant only at high wave numbers, this means that the spanwise fluctuations are the main responsible for dissipating the turbulence kinetic energy. An interesting picture appears then, where turbulence kinetic energy in created mostly on the streamwise fluctuating component of the velocity field, but dissipated mostly by the spanwise fluctuations. At those high wave numbers, the energy content of the spanwise fluctuations is about five times larger than that of the streamwise component near the pipe wall, ruling out the possibility of a bias in the measurements. This observation in Laufer's measurements, indicate that to properly evaluate the viscous dissipation, we must have the spectra of the spanwise velocity fluctuations, because they are the mean responsible for viscous dissipation. Using only the spectrum along the streamwise direction, leads to a much lower value for the total turbulent viscous dissipation. We use Laufer's data set to obtain an averaged spectrum (see supplemental material), and calculate the turbulent part of the viscous dissipation. To evaluate the mean viscous dissipation (from the time-averaged velocity), we use the phenomenological profile derived by Musker\cite{Musker} (see supplemental material). Considering a dimensionless pipe length of $L/2R = 16$, Laufer measured a dimensionless pressure drop of $(P_{inlet}-P_{out}) / (\frac{1}{2}\rho U_{center}^2)=0.16$, where $U_{center}$ is the time--averaged velocity at the center of the pipe. The mean viscous dissipation expressed in the same dimensionless units resulted to be $\rho \overline{\epsilon}_0 L/ (\frac{1}{2}\rho U_{center}^2 V) = 0.058$, which represents only the 36\% of the total pressure drop. 
The results of for the dimensionless pressure drop $\Delta \hat{P} =\Delta P / (\frac{1}{2}\rho U_{center}^2)$, mean viscous dissipation $\hat{\overline{\epsilon}}_0 = \rho \overline{\epsilon}_0 L/ (\frac{1}{2}\rho U_{center}^2 V) $ and turbulent viscous dissipation $\hat{\epsilon'}_0 = \rho \overline{\epsilon}_0 L/ (\frac{1}{2}\rho U_{center}^2 V)$ are presented in table \ref{table0}. Calculating the pressure drop from the addition of $\hat{\overline{\epsilon}}_0 + \hat{\epsilon'}_0$ for all the variants of the averaged spectrum, the error---compared to the experimental value---ranges from 6\% to 20\%.
\begin {table}[h]
 \caption {Dimensionless component of viscous dissipation obtained from the mean velocity profile and averaged spectrum. Arithmetic and geometric (spherical) averages are labeled with a. and g. respectively (see supplemental material for details). Calculations considering an inertial correction are labeled with ``i''.} \label{table0}
\begin{center}
  \begin{tabular}{ | c | c | c | c | c | c |}
    \hline
    $\Delta \hat{P}$ &
    $\hat{\overline{\epsilon}}_0$ & 
    $\hat{\epsilon'}_0$ (a.)&
    $\hat{\epsilon'}_0$ (a.i.)&
    $\hat{\epsilon'}_0$ (g.) &
    $\hat{\epsilon'}_0$ (g.i.) \\ \hline
    0.16 & 0.058 & 0.092 & 0.085 & 0.074 & 0.070 \\ \hline
   \end{tabular}
\end{center}
\end{table}

In conclusion, we have shown that at the incompressibility limit, viscous dissipation cannot increase the temperature of the fluid and the flow's energy can only dissipated into the micro-potential component of the internal energy. We also showed that when introducing a correction to the Kolmogorov scale, as a function of the surface roughness, we can reproduce accurately Nikuradse's \cite{Nikuradse} data for pressure drops in rough pipes. This leads, effectively, to a new correcting term for the energy spectrum in the dissipative regime. The fact that Nikurase's results can be closely reproduced by modifying only the dissipative part of the energy spectrum, is a good indication that small scales in turbulence, play a much important role than previously thought. To validate the use of an isotropic and homogeneous spectrum, in the calculation of the pressure drop, we used the experimental data from Laufer \cite{Laufer}. Our calculations show that the viscous dissipation component associated to the mean velocity profile, only accounts for a modest fraction of the total pressure drop. By using Laufer's experimental spectra, we obtained and averaged isotropic and homogeneous spectrum, which leads to a good agreement between the measured and calculated pressure drop. These results constitute a strong evidence of a direct connection between the pressure drop, viscous dissipation, and the turbulent energy spectrum.

\section*{ACKNOWLEDGEMENTS}
We acknowledge financial support from the Engineering and Physical Sciences Research Council, UK, through the Programme Grant MEMPHIS (grant number EP/K003976/1).

%
%

\appendix

\newcounter{defcounter}
\setcounter{defcounter}{0}
\newenvironment{myequation}{%
\addtocounter{equation}{-1}
\refstepcounter{defcounter}
\renewcommand\theequation{S\thedefcounter}
\begin{equation}}
{\end{equation}}


\section {Conservation laws}
We start our derivation by the most general statement of conservation of a $(n)-$rank tensor $\Phi^{\left(n\right)}$ inside a control a dynamic volume $\Omega\left(t\right)$
\begin{myequation}
\frac{d}{dt}\left(\int_{\Omega\left(t\right)}\boldsymbol{\Phi}^{\left(n\right)} d \Omega\right)
+\int_{d\boldsymbol{\Gamma}\left(t\right)}\textbf{j}_{\Phi}^{\left(n+1\right)}\cdot d\boldsymbol{\Gamma}
= \int_{\Omega\left(t\right)}\textbf{S}_{\Phi}^{\left(n\right)} d\Omega
\label{eqn:conservation1}
\end{myequation}
where $\textbf{j}_{\Phi}^{\left(n+1\right)}$ represents the $\left(n+1\right)-$rank tensor flux of $\boldsymbol{\Phi}^{\left(n\right)}$ through the surface of the control volume $\boldsymbol{\Gamma}\left(t\right)$, and $\textbf{S}_{\Phi}^{\left(n\right)}$ is any source of $\boldsymbol{\Phi}^{\left(n\right)}$. Applying the divergence and Reynolds transport theorem, we arrive at the differential form of the conservation law
\begin{myequation}
\frac{\partial \boldsymbol{\Phi}^{\left(n\right)}}{\partial t} 
+\nabla\cdot\left(\textbf{j}_{\Phi}^{\left(n+1\right)}-\textbf{u}_{\Gamma}\boldsymbol{\Phi}^{\left(n\right)}\right)
= \textbf{S}_{\Phi}^{\left(n\right)} 
\label{eqn:conservation2}
\end{myequation}
with $\textbf{u}_{\Gamma}$ the boundary velocity of the control volume. All conservation laws can be written in this way, provided the proper expressions for total flux $\textbf{j}_{\Phi}^{\left(n\right)}$. In this work, we are concerned with mass, momentum and energy conservation, thus, the fluxes for these quantities are given by: mass $\textbf{j}^{(1)}_{\rho}=\textbf{u}\rho + \textbf{j}_d$, momentum $\textbf{j}^{(2)}_{\textbf{p}}=\textbf{u}\textbf{p} + \stress -\mu\rho\identity$, energy \mbox{$\textbf{j}^{(1)}_e=\textbf{p} e + \textbf{q} + \left(\stress-\mu\rho\identity\right)\cdot\textbf{u}$}. The term $\stress -\mu\rho\identity$ corresponds to the momentum flux due to stresses and chemical species, thus, $\left(\stress -\mu\rho\identity\right)\cdot\textbf{u}$ represents the work rate done by the fluid. By further considering external forces as a source of momentum and energy (if these forces exert work in the system), we arrive at the final set of conservation laws,
\begin{myequation}
\frac{\partial \rho}{\partial t}
+\nabla\cdot\left(\left[\textbf{u}-\textbf{u}_{\Gamma}\right]\rho\right) = -\nabla\cdot\textbf{j}_d
\label{eqn:mass}
\end{myequation}
\begin{myequation}
\frac{\partial \textbf{p} }{\partial t}
+\nabla\cdot\left(\left[\textbf{u}-\textbf{u}_{\Gamma}\right]\textbf{p}\right) = 
-\nabla\cdot\left(\stress-\mu\rho\identity\right) + \Sigma\textbf{f}_{ext}
\label{eqn:momentum}
\end{myequation}
\begin{myequation}
\frac{\partial \left(\rho e\right) }{\partial t}
+\nabla\cdot\left(\left[\textbf{u}-\textbf{u}_{\Gamma}\right] \rho e\right) = -\nabla\cdot\textbf{q} 
-\nabla\cdot\left(\left[\stress-\mu\rho\identity\right]\cdot\textbf{u}\right)
+ \Sigma\textbf{f}_{ext}\cdot\textbf{u}
\label{eqn:energy}
\end{myequation}
Here, $\rho$ is the density, $\mu=\Sigma\left(\mu_i \rho_i \phi_i\right)/\rho$ the averaged chemical potential per unit mass, $\phi$ the volume fraction of each phase or specie, $\textbf{u}$ the velocity of the fluid, $\textbf{j}_d$ a diffusional mass flux (whose form is irrelevant for the current paper), $\textbf{p}=\rho\textbf{u}$ the linear momentum, $\textbf{f}_{ext}$ the external forces, $\stress = P\identity - \deviatoric$ the stress tensor written in terms of the pressure $P$ and the viscous stress tensor $\deviatoric$, $\rho e$ the total energy per unit volume (internal + kinetic) and $\textbf{q}$ the heat flux. The term corresponding to the chemical potential is added for generality purposes, where variation of the chemical potential produce work (e.g. osmotic pressure). For single phase mixtures assumed to be under chemical equilibrium, the chemical potential and density are constant and, therefore, the contribution of this term to the energy and momentum equations is null. However, for two phase flows, $\nabla\left(\mu\rho\right)$ leads to the surface tension force inside of the phase boundary \cite{Badillo}.

An equation for kinetic energy can be obtained by multiplying (inner product) Eq. (\ref{eqn:momentum}) by the velocity and applying mass conservation:
\begin{myequation}
\begin{split}
\frac{\partial \left(\rho e_{k}\right) }{\partial t}
+\nabla\cdot\left(\left[\textbf{u}-\textbf{u}_{\Gamma}\right] \rho e_{k} + 
\left[\stress-\mu\rho\identity \right]\cdot\textbf{u}\right) = \\
\left[\stress-\mu\rho\identity\right]\colon\nabla\textbf{u}
+\Sigma\textbf{f}_{ext}\cdot\textbf{u}
+\frac{\textbf{u}^2}{2}\nabla\cdot\textbf{j}_d
\end{split}
\label{eqn:kinetic}
\end{myequation}
The last term in the RHS of Eq. (\ref{eqn:kinetic}) corresponds to a change in the kinetic energy due to a variation of the density with the local concentration. By subtracting Eq. (\ref{eqn:kinetic}) from Eq. (\ref{eqn:energy}), we obtain an equation for the internal energy:
\begin{myequation}
\begin{split}
\frac{\partial \left(\rho e_{u}\right) }{\partial t}
+\nabla\cdot\left(\left[\textbf{u}-\textbf{u}_{\Gamma}\right] \rho e_{u} \right) = \\
-\nabla\cdot\textbf{q}
-\left[\stress-\mu\rho\identity\right]\colon\nabla\textbf{u} 
-\frac{\textbf{u}^2}{2}\nabla\cdot\textbf{j}_d
\end{split}
\label{eqn:internal}
\end{myequation}
The last term in Eq. (\ref{eqn:internal}) modifies the internal energy due to diffusional fluxes changing the concentration. Writing Eq. (\ref{eqn:internal}) in a non-conservatice manner
\begin{myequation}
\begin{split}
\rho\frac{D^r e_{u}}{D t} = 
-\nabla\cdot\textbf{q}
-\left[\stress-\mu\rho\identity\right]\colon\nabla\textbf{u} \\
+\left(e_{u} 
-\frac{\textbf{u}^2}{2}\right)\nabla\cdot\textbf{j}_d
\end{split}
\label{eqn:internal_non_conservative}
\end{myequation}
where $D^r/Dt=\partial/\partial t + \textbf{u}_r\cdot\nabla$ is the material derivative, based on the relative velocity between the fluid and the control volume boundary $\textbf{u}_r=\textbf{u}-\textbf{u}_{\Gamma}$. Applying the first law of thermodynamics, the change in the internal energy is given $d E_u = \delta Q - \delta W$, where heat and work satisfy $\delta Q \leq TdS$ and $\delta W = PdV - \Sigma \mu_i dN_i$ respectively. At the incompressibility limit, we have $\delta Q = TdS$. By dividing the variation of the internal energy by the mass of the system and applying the material derivative leads to
\begin{myequation}
\frac{D^r e_u}{Dt} \leq T\frac{D^r s}{Dt} - P\frac{D^r \rho^{-1}}{Dt}+\Sigma \mu_i\frac{D^r \left(N_i/m\right)}{Dt}  
\label{eqn:delta_U}
\end{myequation}
the last term can be re-expressed in terms of an effective chemical potential per unit mass as $\left(\overline{\mu}/\rho \right)D^r\rho/Dt$. Hence, the change in the internal energy is
\begin{myequation}
\frac{D^r e_u}{Dt} \leq T\frac{D^r s}{Dt} + \frac{1}{\rho^2}\left(P+\overline{\mu}\rho\right)\frac{D^r\rho}{Dt}  
\label{eqn:delta_U1}
\end{myequation}
and substituting Eq. (\ref{eqn:delta_U1}) into Eq. (\ref{eqn:internal_non_conservative}) leads to the inequality for the entropy
\begin{myequation}
\begin{split}
\rho T\frac{D^r s}{Dt} \geq 
-\nabla\cdot\textbf{q}
-\left[\stress-\mu\rho\identity\right]\colon\nabla\textbf{u} \\
-\frac{1}{\rho^2}\left(P+\overline{\mu}\rho\right)\frac{D^r\rho}{Dt}
+\left(e_{u} -\frac{\textbf{u}^2}{2}\right)\nabla\cdot\textbf{j}_d
\end{split}
\label{eqn:entropy}
\end{myequation}
%
%
%
%
%
%
%
By replacing the thermodynamic definition of the enthalpy per unit mass $h=e_u+P\rho^{-1}$ into Eq. (\ref{eqn:internal}) we arrive at
\begin{myequation}
\begin{split}
\frac{\partial \left(\rho h\right) }{\partial t}
+\nabla\cdot\left(\left[\textbf{u}-\textbf{u}_{\Gamma}\right] \rho h + \textbf{q}\right) = 
\frac{\partial P }{\partial t}
+\nabla\cdot\left(\left[\textbf{u}-\textbf{u}_{\Gamma}\right] P \right)\\
-\left[\stress-\mu\rho\identity\right]\colon\nabla\textbf{u} 
-\frac{\textbf{u}^2}{2}\nabla\cdot\textbf{j}_d
\end{split}
\label{eqn:enthalpy}
\end{myequation}
Replacing the definition of the stress tensor ${\stress = P \identity - \deviatoric}$ and considering a pure single phase incompressible system with a fixed control volume ($\textbf{u}_{\Gamma}=0$), the equations for kinetic energy, internal energy, entropy and enthalpy respectively reduce to
\begin{myequation}
\frac{\partial \left(\rho e_{k}\right) }{\partial t}
+\nabla\cdot\left(\textbf{u}\rho e_{k} + \textbf{u}P - \deviatoric\cdot\textbf{u}  \right) =
 -\deviatoric\colon\nabla\textbf{u}
+\Sigma\textbf{f}_{ext}\cdot\textbf{u}
\label{eqn:kinetic2}
\end{myequation}
\begin{myequation}
\frac{\partial \left(\rho e_{u}\right) }{\partial t}
+\nabla\cdot\left(\textbf{u}\rho e_{u} \right) = 
-\nabla\cdot\textbf{q}+\deviatoric\colon\nabla\textbf{u} 
\label{eqn:internal2}
\end{myequation}
\begin{myequation}
\rho T\frac{D s}{Dt} = 
-\nabla\cdot\textbf{q}
+\deviatoric\colon\nabla\textbf{u}
\label{eqn:entropy2}
\end{myequation}
\begin{myequation}
\begin{split}
\frac{\partial \left(\rho h\right) }{\partial t}
+\nabla\cdot\left(\textbf{u}\rho h \right) = -\nabla\cdot\textbf{q} + \\
\frac{\partial P }{\partial t} 
+\nabla\cdot\left(\textbf{u}P \right)+\deviatoric\colon\nabla\textbf{u} 
\end{split}
\label{eqn:enthalpy2}
\end{myequation}
 It is important to remark that Eq. (\ref{eqn:entropy2}), is only valid at the incompressibility limit. For compressible (real) flows, this equation changes to an inequality, following the second law of thermodynamics.

\section{Pressure drop}

We obtain the first expression for the pressure drop, from the time-averaged momentum equation
\begin{myequation}
\rho \nabla \cdot \left( \overline{\textbf{u}} \overline{\textbf{u}} \right ) = -\nabla \overline{P}  -  \nabla\cdot (\rho\left\langle \textbf{u}' \textbf{u}' \right\rangle - \mu\nabla\overline{\textbf{u}} -\mu\nabla\overline{\textbf{u}}^T)
\end{myequation} 
with $ \overline{\textbf{u}} = \left\langle \textbf{u} \right\rangle$ and $ \overline{P} = \left\langle P \right\rangle$ time-averaged quantities. To arrive to this equation, we use the Reynolds decomposition for velocity $\textbf{u}\left(\textbf{x},t\right)=\overline{\textbf{u}}\left(\textbf{x}\right)+\textbf{u}'\left(\textbf{x},t\right)$ and pressure $P \left(\textbf{x},t\right) = \overline{P}\left(\textbf{x}\right)  + P'\left(\textbf{x},t\right) $. For fully developed flow, 
$\rho \nabla \cdot \left( \overline{\textbf{u}} \overline{\textbf{u}} \right ) = 0$ and, hence, $\nabla\overline{P} =  -  \nabla\cdot (\rho\left\langle \textbf{u}' \textbf{u}' \right\rangle - \mu\nabla\overline{\textbf{u}} -\mu\nabla\overline{\textbf{u}}^T)$. Because the Reynolds stresses $\rho\left\langle\textbf{u}' \textbf{u}' \right\rangle$ are the same at the inlet and outlet due to the fully developed flow assumption, the surface integral of $\nabla\cdot \left(\rho\left\langle \textbf{u}' \textbf{u}' \right\rangle\right)$  vanishes, from which we arrive to 
\begin{myequation}
\Delta P = -\frac{2L}{R}\rho u_{\tau}^2
\label{eqn:pressureDrop1}
\end{myequation}
with $u_{\tau}=\sqrt{\tau_w/ \rho}$ the friction velocity,  $\tau_w$ the constant wall shear stress, $L$ the pipe length, and $R$ the pipe radius. For the second expression of the pressure drop, we integrate over the surface the time-averaged conservation equation of kinetic energy
\begin{myequation}
\int_{A}\left\langle \textbf{u}\rho e_k  + \textbf{u}P - \deviatoric\cdot\text{u}\right\rangle \cdot d\text{A} = 
-\int_{\Omega}\left\langle \deviatoric\colon\nabla\textbf{u} \right\rangle d\Omega 
\label{eqn:kinetic3}
\end{myequation}
Since for fully developed turbulent pipe flow fluctuating quantities are translation invariant, we have that ${ \int_{A}\left\langle \textbf{u}\rho e_k\right\rangle \cdot d\textbf{A}=0}$ and  ${\int_{A} \left\langle\deviatoric \cdot \textbf{u}\right\rangle \cdot d\textbf{A}}=0$. This means that the average flow rate of kinetic energy and the work done by viscous stresses at inlet and outlet are equal. By applying the Reynolds decomposition to the pressure and velocity, we arrive to
\begin{myequation}
\int_A \left(\left\langle P' \textbf{u}' \right\rangle + \overline{P}\overline{\textbf{u}}\right) \cdot d\textbf{A} = 
-\int_{\Omega}\left\langle \deviatoric\colon\nabla\textbf{u}\right\rangle d\Omega 
\label{eqn:kinetic8}
\end{myequation}
Since the velocity is zero at walls, the surface integrals reduce only to the inlet and outlet. Invoking translational invariance along the stream wise direction, the integration of $\left\langle \widehat{P}\widehat{\textbf{u}}\right\rangle$ over the surface vanishes and, hence
\begin{myequation}
\Delta P =- \frac{1}{VA}\int_{\Omega}\left\langle \deviatoric\colon\nabla\textbf{u}\right\rangle d\Omega 
\label{eqn:pressureDrop2}
\end{myequation}     
where $VA=\int_A \textbf{u}\cdot d\textbf{A}$ is the volumetric flow rate. For fully developed laminar pipe flow the velocity field has only one component in the axial direction $\textbf{u}=\left(0,0,u_z\right)$, then the tensor velocity gradient in cylindrical coordinates is given by
\begin{myequation}
\nabla \textbf{u} =
\left(
\begin{array}{ccc}
0 & 0 & \frac{\partial u_z}{\partial r} \\
0 & 0 & 0 \\
0 & 0 & 0\\

\end{array}
\right)
\end{myequation}
Thus, the local dissipation is \mbox{$\deviatoric\colon\nabla\textbf{u}=\mu\left(\frac{\partial u_z}{\partial r}\right)^2 $}. Replacing the parabolic velocity profile \mbox{ $u_z\left(r\right) = 2V\left(1-\left(r/R\right)^2\right)$} into Eq. (\ref{eqn:pressureDrop2}) leads to
\begin{myequation}
\Delta P = -\frac{8\pi\mu V L}{A}
\label{eqn:pressureDrop3}
\end{myequation}     
The same result is obtained when using the parabolic velocity profile to calculate the wall shear stress in Eq. (\ref{eqn:pressureDrop1})
%
%
\section{Fitting Molecular Dynamics data}\label{MDdata}
\begin{figure}[htbp]
	\centering
	\includegraphics[width=8.5cm]{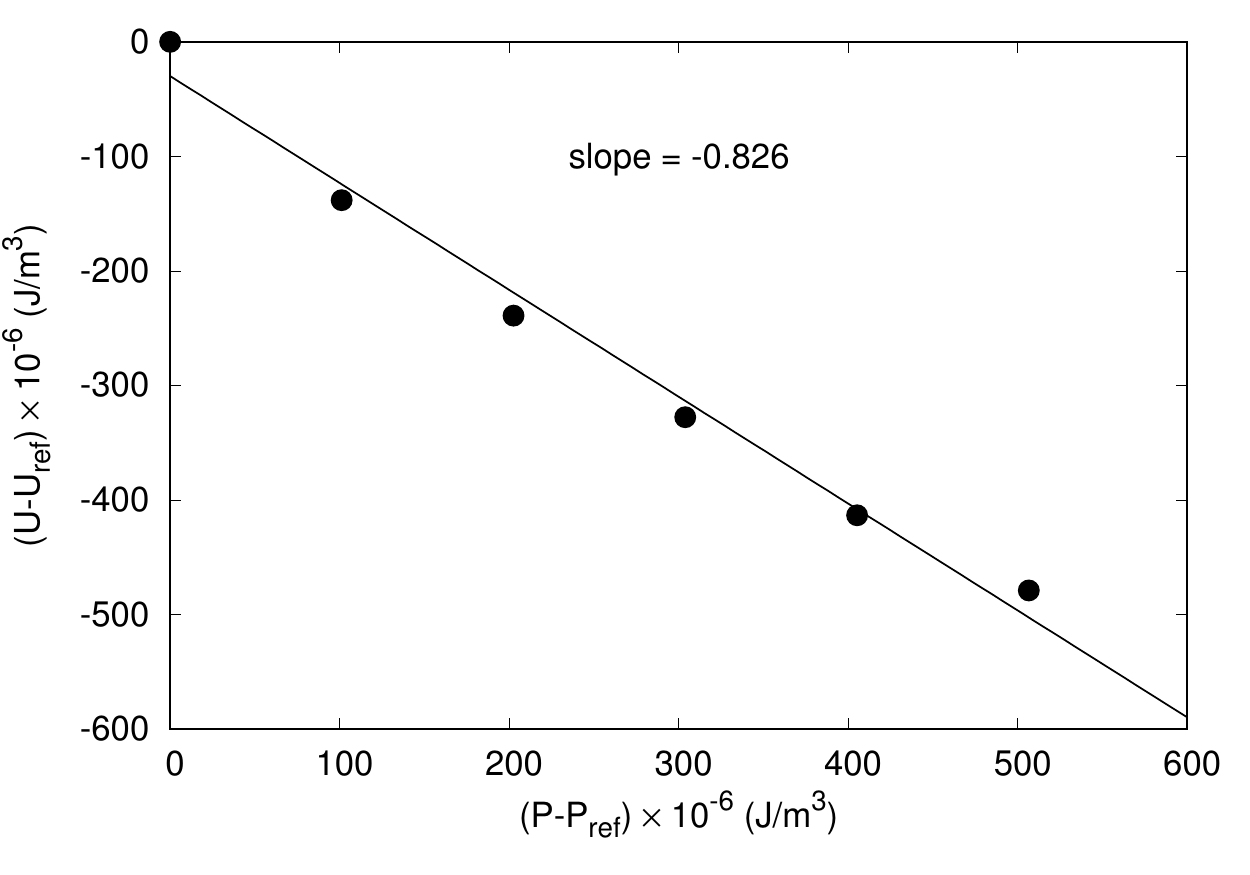}
	\caption{MD results for the variation of internal energy with pressure. The data were obtained from Ref. 
\cite {Mahoney}. Reference values for pressure and internal energy were taken at 1 atmosphere. To express the internal energy in units of energy per unit volume, we used a molar mass for water of 18.015 g/mol; the density was taken from the MD data.}
	\label{fig:image1}
\end{figure}

Molecular Dynamics simulations, are the only tools that can allows us to quantify simultaneously the effects of pressure on the internal energy and the local range order of a liquid. It is well known that changes in the molecular arrangement of a liquid has an associated change in the internal energy (due to changes in the energy of each molecular bond). In the case of water at high pressures, it is possible to modify the angle of the hydrogen bonds, while keeping and almost unchanged density. This could allows us to modify greatly the internal energy with pressure, while keeping a constant density. As mentioned in the main body of the article, for a strictly incompressible fluid, the derivative of the internal energy respect to pressure is equal to $d e_u /dP =-1$. We used the the MD data from Mahoney and Jorgensen \cite{Mahoney} to estimate this derivative. Figure (\ref{fig:image1}) shows the MD data along with a linear fit, whose slope resulted to be $-0.826$. This indicates a slightly compressible behavior of the MD fluid simulated by these authors. Having a slope close to $-1$ means that a large fraction of the internal energy that is borrowed, when pressure is increased, is paid back following a reversible path. Flow compressibility leads to irreversible changes of the internal energy, with the consequent generation of sensible heat.

%
\section{Derivation of the second equation for the pressure drop from conventional time averaged equations.}\label{popesEquations}
In spite that Eq. (\ref{eqn:pressureDrop2}) appears to be in contrast with Townsend's hypothesis \cite{Townsend}, where the production of turbulence kinetic energy balances the viscous dissipation near the wall, we will see that the contradiction is only apparent.  Pope \cite{Pope} presents in his book separate equations (5.131 and 5.132) for conservation of the mean and turbulence kinetic energy. Using our nomenclature, these equations read
\begin{myequation}
\frac{\partial \overline{e}_k}{\partial t} + \overline{\textbf{u}} \cdot \nabla \overline{e}_k + \nabla \cdot \overline{\textbf{T}} = -\Production - \overline{\Dissip}
\label{eqn:pope1}
\end{myequation}
\begin{myequation}
\frac{\partial {e}_{TKE}}{\partial t} + \overline{\textbf{u}} \cdot \nabla {e}_{TKE} + \nabla \cdot {\textbf{T'}} = \Production - \Dissip'
\label{eqn:pope2}
\end{myequation}
with $\overline{\textbf{T}}= \overline{\textbf{u}}\cdot \langle \textbf{u}'\textbf{u}' \rangle + \overline{\textbf{u}}\overline{P} / \rho -2\nu \langle \overline{\textbf{u}}\cdot \nabla \overline{\textbf{u}} \rangle $, $\textbf{T}' = 0.5\langle \textbf{u}'\cdot \textbf{u}'\textbf{u}' \rangle+ \textbf{u}' P' / \rho - 2 \nu \langle \textbf{u}'\textbf{u}' \rangle$, the production of kinetic energy \mbox{$\Production = - \langle \textbf{u}'\textbf{u}' \rangle: \nabla \overline{\textbf{u}}$}, mean dissipation $\overline{\Dissip} = \langle\deviatoric\rangle:\left( \nabla \overline{\textbf{u}} + \nabla \overline{\textbf{u}}^T \right) / \rho $ and turbulent dissipation \mbox{$\Dissip' = \left\langle \deviatoric' :\left( \nabla \textbf{u}' + \nabla \textbf{u}'^T \right) / \rho\right\rangle $}. A detailed analysis of conservation of kinetic energy, Eq. (\ref{eqn:kinetic}), shows that the local instantanous viscous dissipation is given by  $\Dissip = \deviatoric : \nabla \textbf{u} / \rho $. Because of the symmetry of the stress tensor, the dissipation can also be expressed in terms of the strain rate tensor as  $\Dissip = \deviatoric : \left(\nabla \textbf{u}  + \nabla \textbf{u}^ T \right)/2/ \rho $. In our article, we derived the equation for the pressure drop assuming conservation of (total) kinetic energy in a pipe under fully developed flow conditions. Applying this flow regime to Eqs. (\ref{eqn:pope1} and \ref{eqn:pope2}) and integrating over the pipe volume, leads to
\begin{myequation}
\int_A \overline{\textbf{u}} \overline{P} \cdot d\textbf{A}= -\int_\Omega \Production d\Omega - \int_\Omega \overline{\Dissip} d\Omega
\label{eqn:pope3}
\end{myequation}
\begin{myequation}
0 = \int_\Omega\Production d\Omega - \int_\Omega \Dissip' d\Omega
\label{eqn:pope4}
\end{myequation}
Conservation of kinetic energy requires the fulfillment of both equations simultaneously. This means that under the current flow conditions, the global production of turbulence kinetic energy balances exactly the global turbulent component of the viscous dissipation, in agreement with Townsend's hypothesis. However, by replacing Eq. (\ref{eqn:pope4}) into (\ref{eqn:pope3}), we arrive to Eq. (\ref{eqn:pressureDrop2}). Equation (\ref{eqn:pope3}) tells us that the pressure gradient is in charge of creating turbulence kinetic energy and overcome the mean viscous dissipation, but once kinetic energy is created, it is dissipated into internal energy requiring more energy--in the form of pressure--to sustain the steady state and fully developed flow conditions.

\section{Determining the dissipation scale}
To determine the dissipation scale $\tilde{\epsilon}$, we can make use of the friction factor defined in the main body of the article
\begin{myequation}
f = \frac{2 D_H}{V^3}\frac{1}{\Omega} \int_{\Omega}\overline{\epsilon}_0 d\Omega
   +\frac{2 D_H}{V^3}\frac{1}{\Omega} \int_{\Omega}\epsilon'_0 d\Omega
   \label{eqn:friction}
\end{myequation}  

\noindent As discussed in the text, the component associated to $\overline{\epsilon}_0$ vanishes for high Reynolds numbers. Thus, to have a constant friction factor as $\Rey\rightarrow\infty$, the second term must be constant. Defining the dissipation scale as
\begin{myequation}
  \tilde{\epsilon}=\frac{1}{\Omega} \int_{\Omega}\epsilon'_0 d\Omega
\end{myequation}
we arrive at once to
\begin{myequation}
\tilde{\epsilon} = C_{\epsilon}\frac{V^3}{D_H}
\label{eqn:dissipation_scale}
\end{myequation} 
\noindent with $C_{\epsilon}$ a dimensionless constant. This dissipation scale is consistent with the upper bound derived by Doering and Constantin\cite{Doering} in shear driven turbulence. It is interesting to note that this dissipation scale is only valid for the Strickler regime ($\Rey \rightarrow \infty$). The fact that the friction factor changes from the Blausius regime (dominated by a power law with an exponent equal to $-1/4$), to the Strickler regime where the friction factor is constant and independent from the Reynolds number, indicates that there is a flow transition. Just as the transition from laminar to fully turbulent occurs in a Reynolds number interval, this second transition also happens in a certain interval, which in this case, depends on the surface roughness. Because our main hypothesis is that the friction factor depends exclusively on the viscous dissipation, the different plateaus observed by Nikuradse\cite{Nikuradse} for the friction factor at high Reynolds number, indicates that the flow transition should happen only at high wave numbers. In other words, we should observe modifications to the energy spectrum only in the dissipative zone.

\section{Determining the scaling function $\psi\left(R/\eta,\Rey, \lambda/R\right)$} \label{phiFunction}

\begin{figure}[htbp]
	\centering
		\includegraphics[width=8cm]{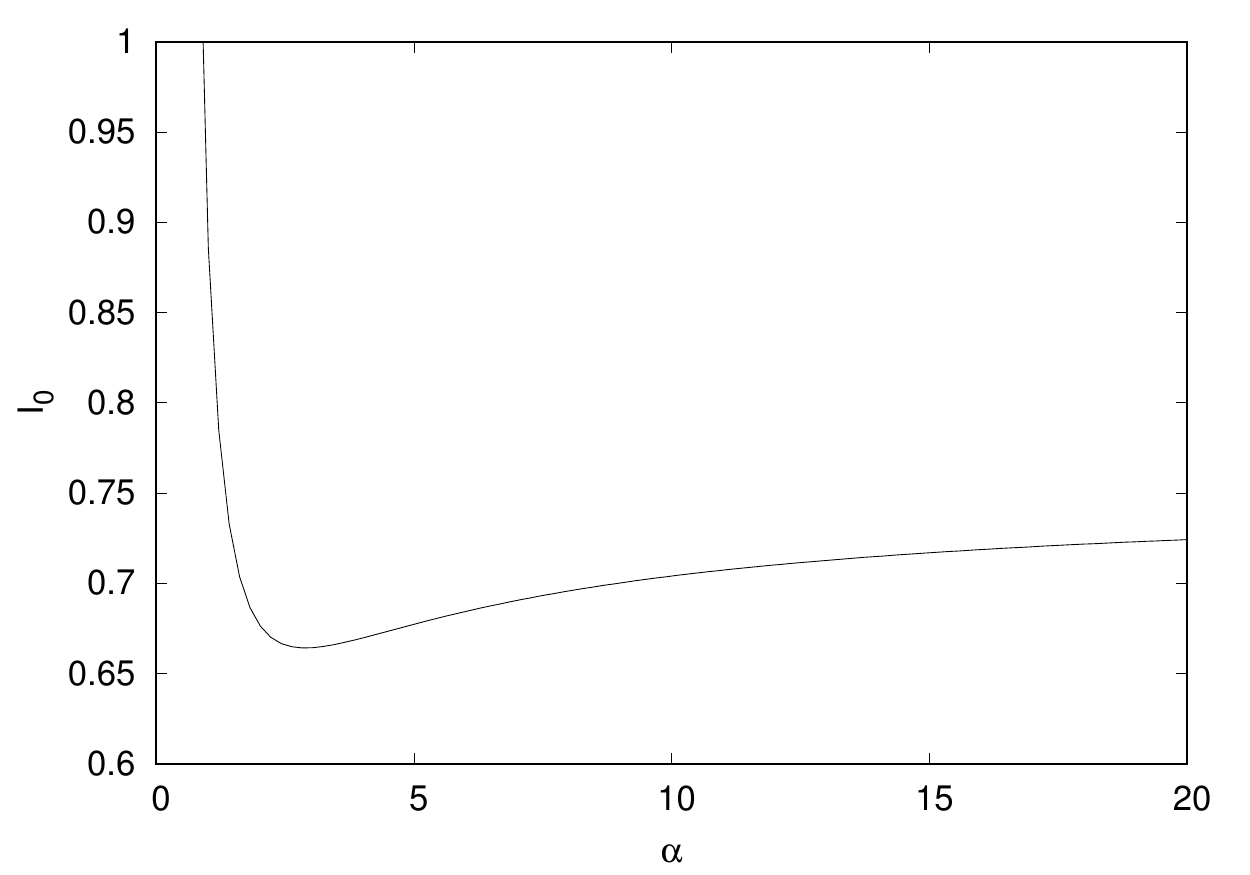} 		
	\caption{Variation of the integral $I_0$ respect to $\alpha$ (the power exponent in correcting term in the dissipation controlled regime). }
	\label{fig:image2}
\end{figure}

Using the phenomenological spectrum, the turbulent component of the fraction factor is written as
\begin{myequation}
f_T=\frac{64}{\Rey} \frac{15A_0 \tilde{\epsilon}^{2/3}\Omega}{8\pi V^2}\int_0^\infty q^{1/3}\exp\left(-\left[ g\left(\eta,\lambda/D_H\right) q\right]^{\alpha} \right) dq
\label{eqn:fl_friction}
\end{myequation}
\noindent with $g$ being a function having units of length. Following the variable change $y = g\left(\eta,\lambda/D_H\right) q$, the integral is expressed as
\begin{myequation}
I = g\left(\eta,\lambda/D_H\right)^{-4/3}\int_0^\infty y^{1/3}\exp(-y^{\alpha})dy
\end{myequation}
Evaluating the integral leads to
\begin{myequation}
I = g\left(\eta,\lambda/D_H\right)^{-4/3} I_0
\end{myequation}
with
\begin{myequation}
 I_0=\frac{1}{\alpha}\Gamma\left( \frac{4}{3 \alpha} \right)
\end{myequation}
where $\Gamma$ is the gamma function. Minimizing $I_0$ respect to $\alpha$ is equivalent to minimize the dissipation rate. Figure (\ref{fig:image2}) presents the variation of $I_0$ as a function of $\alpha$, with a minimum value located at $\alpha = 2.8883$.
Making use of the dissipation scale Eq. (\ref{eqn:dissipation_scale}), we can express the turbulent component of the friction factor as
\begin{myequation}
f_T=30 D_H^2 A_0 I_0 \left(\frac{C_{\epsilon}}{D_H}\right)^{2/3}\frac{1}{\Rey^{1/4}}\left(\frac{g ^{-4/3}}{\Rey^{3/4}}\right)
\label{eqn:fl_friction1}
\end{myequation}
\noindent Comparing this expression to the one provided by Goldelfeld \cite{Nigel} $f_T = G / \Rey^{1/4}$, we arrive at
\begin{myequation}
G\left(\Rey^{3/4}\left(\lambda/R\right)\right) = 30 D_H^2  A_0 I_0 \left(\frac{C_{\epsilon}}{D_H}\right)^{2/3} \left(\frac{g ^{-4/3}}{\Rey^{3/4}}\right)
\label{eqn:res0}
\end{myequation}
Introducing Kolmogorov scale, $\eta=\nu^{3/4}\tilde{\epsilon}^{-1/4}$, we obtain
\begin{myequation}
G\left(\Rey^{3/4}\left(\lambda/R\right)\right) = \gamma \eta g^{-4/3}
\label{eqn:res1}
\end{myequation}

\noindent with $\gamma=30 D_H  A_0 I_0 \left(\frac{C_{\epsilon}}{D_H}\right)^{2/3}C_{\epsilon}^{1/4}$ expressed in units of ${\rm length}^{1/3}$. We now propose the following form for $G$
\begin{myequation}
G\left(\Rey^{3/4}\left(\lambda/R\right)\right) = A_2\left(1-\phi_2(z)\right) + A_3\phi_2(z)\Rey^{1/4}\left(\frac{\lambda}{R}\right)^{1/3}
\label{eqn:res2}
\end{myequation}

\noindent where $z = \Rey^{3/4}(\lambda/R)$. 
The blending function $\phi_2(z)$ provides a smooth transition between the Blasius and Strickles regimes. Following Bobby and Joseph \cite{Bobby2}, we have chosen
\begin{myequation}
 \phi_2(z)=1-\frac{1}{1 + \left(z/z_0\right)^2}
\end{myequation}
By equating Eqs. (\ref{eqn:res1},\ref{eqn:res2}), we finally arrive at
\begin{myequation}
g = \eta\left(\frac{\gamma^{3/4}}{\eta^{1/4}\left[A_2\left(1-\phi_2\right)+A_3\phi_2\Rey^{1/4}(\lambda/R)^{1/3}\right]^{3/4}}\right)
\end{myequation}
\noindent Since $\gamma^{3/4}$ has units of $length^{1/4}$, the large parenthesis is dimensionless. Thus, by choosing the dimensionless constant $\beta = \left(30 A_0 I_0 C_{\epsilon}^{11/12}\right)^{3/4}$ we identify 
\begin{myequation}
\psi = \frac{\left(2R/\eta\right)^{1/4}}{\left[A_2\left(1-\phi_2\right)+A_3\phi_2\Rey^{1/4}(\lambda/R)^{1/3}\right]^{3/4}}
\end{myequation}
Equation \ref{eqn:res0} implies that the turbulent component of the friction factor is given by
\begin{myequation}
f_T = A_2\left(1-\phi_2(z)\right)\Rey^{-1/4} + A_3\phi_2(z)\left(\frac{\lambda}{R}\right)^{1/3}
\end{myequation}
This equation arises from the direct blending of the Blasius and Strickles regimes (through $\phi_2$), with no further theoretical support other than experimental observation. Nonetheless, and despite its phenomenological character, this form of the turbulent friction factor provide us with the needed information to explain Nikuradse's results in terms of an averaged Kolmogorov length scale that varies with the surface roughness. By including the laminar contribution, the total friction factor reads
\begin{myequation}
f = \frac{64}{Re} \left(1-\phi_1 \right) + A_2\left(1-\phi_2\right)\phi_1\Rey^{-1/4} + A_3\phi_2 \phi_1 \left(\frac{\lambda}{R}\right)^{1/3}
\end{myequation}
This is the function used when fitting Nikuradse's results.

\section{The idea behind the averaged spectrum}
To investigate the connection between the friction factor, with the turbulent energy spectrum, we introduce the tensor correlation $\corr_{ijk}\left(\textbf{x},\delta\textbf{x}_k\right)=\left\langle u'_i\left(\textbf{x},t\right) u'_j\left(\textbf{x}+\delta\textbf{x}_k,t\right) \right\rangle$, where $i$ and $j$ represent the components of the fluctuating velocity field and $k$ the direction in which the correlation is being calculated. Since $\delta\textbf{x}_k$ can be chosen independently from $\textbf{x}$, $\corr_{ijk}$ is a 3-rank tensor, with eighteen independent components. The position and orientation-dependent energy spectrum is directly obtained by applying the Fourier transform to the tensor correlation, this is, \mbox{$\fourier_{ijk}\left(\textbf{x},\textbf{q}_k\right) = \int_{-\infty}^{\infty} e^{-2\pi i\textbf{q}_k\cdot\delta\textbf{x}_k }\corr_{ijk}\left(\textbf{x},\delta\textbf{x}_k\right) d\delta\textbf{x}_k$}. Since no assumption has been made in the definition of the tensor correlation, this definition of the energy spectrum is exact. Using a Taylor expansion to express the tensor correlation in terms of a position and orientation-independent correlation plus corrections, we have 
$\corr_{ijk}=\corr_0+\left.\partial_{\textbf{x}}\corr_{ijk}\right|_0 * \Delta\textbf{x}+\left.\partial_{\delta\textbf{x}}\corr_{ijk}\right|_0 *\Delta\delta\textbf{x}_k+\cdots$, where $(*)$ indicates convolution (see section \ref{taylorExpansion}). Here, $\corr_0$ is the homogeneous tensor correlation under rotational invariance. 

We assume now that the Taylor expansion is convergent and that first order corrections are the leading terms. The first and second terms represent the departure from homogeneity and isotropy, respectively. Thus, the homogeneous isotropic tensor $\corr_0$ can be interpreted as a spherical average followed by space average of $\corr_{ijk}$. With this, the energy spectrum can be finally written as $\fourier_{ijk}\left(\textbf{x},\textbf{q}_k\right) = \hat{R}_0\left(q\right)+\delta\fourier_t\left(\textbf{x},q\right)\Delta\textbf{x}+\delta\fourier_r\left(\delta\textbf{x},\delta\textbf{q}_k\right)\Delta\delta\textbf{x}+\cdots$. This enables us to calculate each component of the viscous dissipation easily, that is, $\left\langle\left(\partial_i u'_j\right)^2\right\rangle = \int_{-\infty}^{\infty}\textbf{q}_i^2\fourier_{jji}\left(\textbf{x},\delta\textbf{x}_i=0\right)d\textbf{q}_i$ and $\left\langle\partial_i u'_j\partial_j u'_i \right\rangle = \int_{-\infty}^{\infty}\textbf{q}_{\left(ij\right)}^2\fourier_{ji\left(ij\right)}\left(\textbf{x},\delta\textbf{x}_{\left(ij\right)}=0\right)d\textbf{q}_{\left(ij\right)}$. These expressions are exact and the subscripts $\left(ij\right)$ indicate a spectrum obtained along $\delta\textbf{x}_k=\delta\textbf{x}_i+\delta\textbf{x}_j$. Hence, the connection between the energy spectrum and the pressure drop is directly realized through the Darcy-Weisbach equation (defined in the main body) and Eq. (\ref{eqn:friction}). 
Following Taylor's \cite{Taylor} identities for isotropic turbulence, we have that the fluctuating component of the dissipation rate can be approximated by \mbox{$\epsilon'_0\approx 15\nu\int_0^\infty q^2 \hat{R}_0 dq + \mathcal{O}\left(\Delta\textbf{x}+\Delta\delta\textbf{x}_k\right)$}. 

Our hypothesis is that errors introduced in the calculation of the total turbulent viscous dissipations, are of second order when using an averaged spectrum. This was confirmed by using the experimental data from Laufer \cite{Laufer}.  
\section{Spherically and spatially averaged spectrum} \label{averagedSpectrum}
In this letter, we propose that the pressure drop can be determined from an spherically and spatially averaged spectrum. We calculate the spherical average by using an arithmetic and geometric average, and compare its influence on the calculation fo the pressure drop. These two spherical averages are given by 
\begin{myequation}
\overline{E}_s\left( q,r \right) = \frac{1}{3}\left(E_{u'u'}\left( q,r \right)  + E_{v'v'}\left( q,r \right)  + E_{w'w'}\left( q,r \right) \right) 
\end{myequation}
\begin{myequation}
\overline{E}_s\left( q,r \right) = \left(E_{u'u'}\left( q,r \right) E_{v'v'}\left( q,r \right) E_{w'w'}\left( q,r \right) \right)^\frac{1}{3}
\end{myequation}
The geometric average is the only one that allows us to express the spherically averaged correction in the dissipative regime $C_d$ with the same functional form as the individual spectra. Thus, for the geometrically averaged spectrum, the dissipation correction is expressed as $e^{- \overline{B} q^\alpha}$, with $\overline{B} = \frac{B_{u'u'} + B_{v'v'} + B_{w'w'}} {3}$. Since the constant $D$ in the fitted spectrum is different for each velocity component, the correction to the inertial regime $\left( 1 + \frac{D}{q^{\alpha}} \right)^{-C}$ cannot be expressed in terms of an averaged exponent $\overline{C}$. However, and as seen in the calculation of the total dissipation rate (see results in the main text), the effects of the inertial regime are negligible.
The spatial average was carried out by using a linear interpolation of the spherically averaged spectra between the measured radial positions
\begin{myequation}
\begin{split}
\langle E\left( q\right) \rangle  = \frac{1}{\pi R^2} \int_0^R \overline{E}_s\left( q,r \right) 2 \pi r dr \\
\approx \frac{2}{R_{max}^2}
 \sum_{i = 1}^{n}  \int_{R_i}^{R_{i+1}} \overline{E}_s\left( q,r \right) r dr
\end{split} 
\end{myequation}
where $R_{max}$ is the closest (measured) position to the wall.

\section{Fitting Individual spectra} \label{individualSpectra}
The function for fitting each individual spectrum is given by

\begin{myequation}
E\left( q \right) = A q^{-5/3} e^{-B q^{\alpha}} \left( 1 + \frac{D}{q^{\alpha}} \right)^{-C}
\end{myequation}
with the parenthesis and exponential term constituting the corrections to the inertial and dissipation regimes respectively. Hence, the departure from the $5/3$ law might be interpreted as the combined effect introduced by the inertial and dissipation regimes. 
\begin {table}[h]
 \caption {Fitted coefficients for individual spectra. $(*)$ Interpolated value (experimental data is not available).} \label{table1}
\begin{center}
  \begin{tabular}{ | c | c | c | c | c | c | c |}
   
    \hline
    Stress &
    $r/R$  &  $A^*$      &     $B$      &      $C$      &      $D$   &  $ \eta \left(cm\right) $ \\ \hline 
    $u'u'$ &0		&7.147	&2.866E-4	&0.4896	&0.01512	&0.0368 \\ \hline
    $u'u'$ &0.309	&7.207	&1.272E-4	&0.4017   &0.01198	&0.0330 \\ \hline
    $u'u'$ &0.72		&7.532	&1.473E-5	&0.2669	&0.03245	&0.0237 \\ \hline
    $u'u'$ &0.926	&8.291	&2.433E-5	&0.3170	&0.09888	&0.0208 \\ \hline
    $u'u'$ &0.9918	&9.481	&2.144E-6	&0.2530	&100	.0	&0.0117 \\ \hline
    
    $v'v'$ &0		&7.621	&7.562E-5	&0.6333	&0.07428	&0.0280 \\ \hline
    $v'v'$ &0.309	&7.833	&4.831E-5	&0.6255	&0.07369	&0.0252 \\ \hline
    $v'v' (*)$ &0.72	&8.411	&7.093E-6	&0.4176	&2.23390	&$---$	\\ \hline
    $v'v'$ &0.926	&9.382	&1.303E-5	&0.4506	&20.4591	&0.0148\\ \hline
    $v'v' (*)$ &0.9918	&10.771	&1.201E-6	&0.3429	&22033.4	&$---$	\\ \hline
    
    $w'w'$ &0		&7.684	&8.130E-5	&0.6434	&0.07824	&0.0278 \\ \hline
    $w'w'$ &0.309	&8.030	&9.208E-5	&0.6483	&0.06212	&0.0259 \\ \hline
    $w'w'(*)$ &0.72	&8.610	&1.726E-5	&0.4030	&0.97417	&$---$	\\ \hline
    $w'w'$ &0.926	&9.600	&3.415E-5	&0.3940	&6.5343	&0.0157 \\ \hline
    $w'w'(*)$ &0.9918	&11.056	&3.209E-6	&0.2852	&6695.1	&$---$	\\ \hline
    
  \end{tabular}
\end{center}
\end{table}

\begin{figure}[!htb]
        \includegraphics[width=8cm]{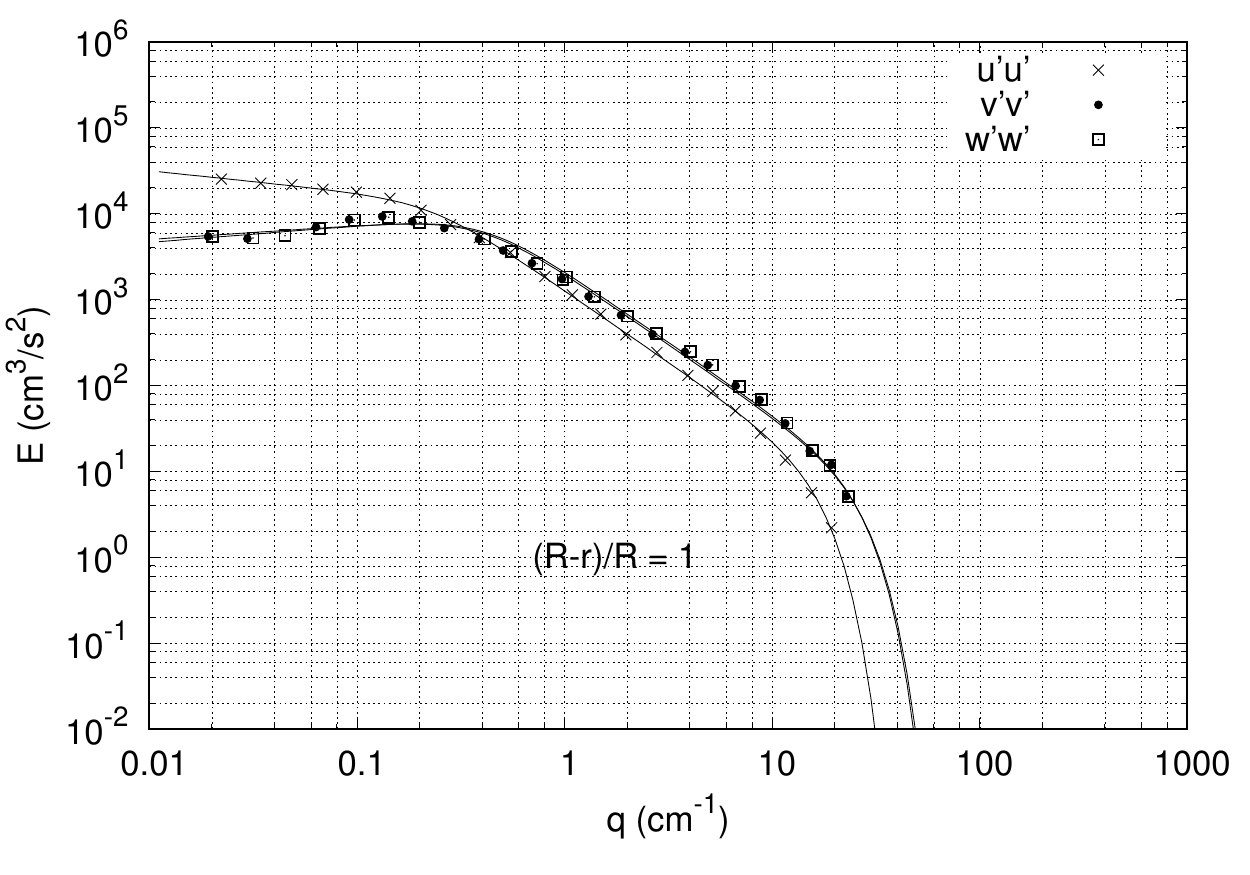}
	\caption{Fitted spectra at the pipe center}
	\label{fig:uSpectraCenter}
\end{figure}

\begin{figure}[!htb]
        \includegraphics[width=8cm]{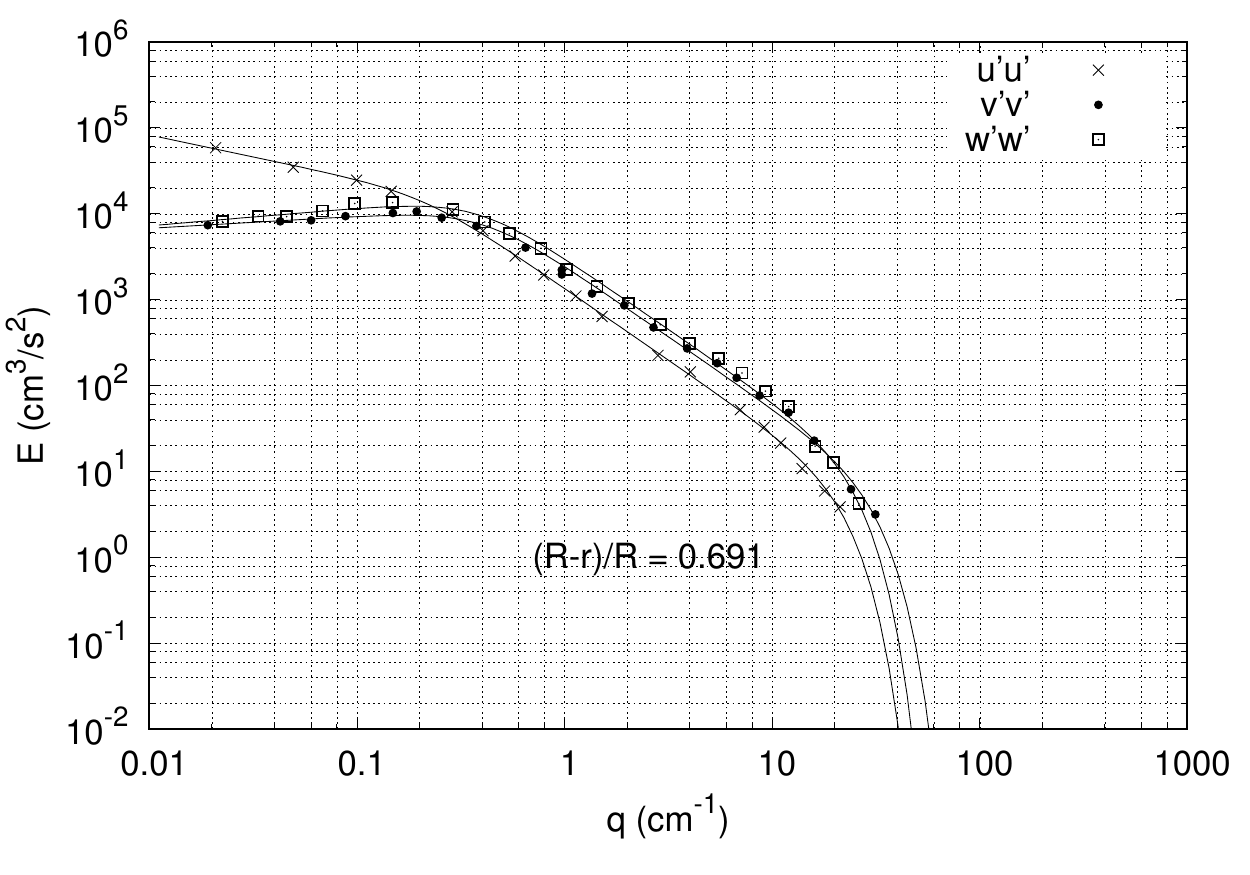}
	\caption{Fitted spectra at $\left(R -r \right)/R = 0.691$}
	\label{fig:uSpectraR1}
\end{figure}

\begin{figure}[!htb]
        \includegraphics[width=8cm]{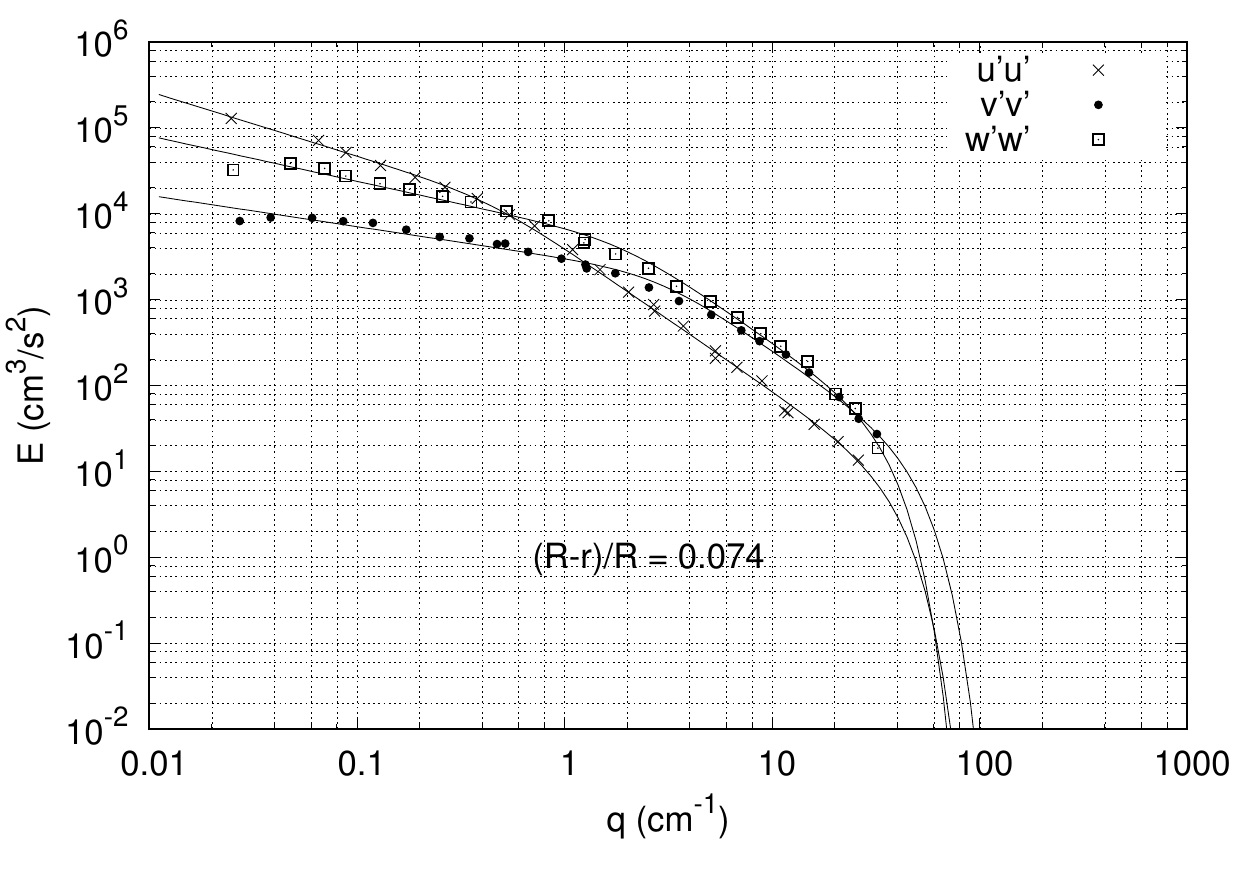}
	\caption{Fitted spectra at $\left(R -r \right)/R = 0.074$}
	\label{fig:uSpectraR2}
\end{figure}

\noindent Since the experimental results from Laufer are presented in logarithmic scale, we apply a logarithmic fit
\begin{myequation}
\ln \left \{ E\left( q \right) \right\} = A^* - \frac{5}{3} q^* -B e^{\alpha q^*}  -C \ln \left( 1 + \frac{D}{e^{\alpha q^*}} \right)
\end{myequation}
where $A^* = \ln \left(A \right)$, $q^* = \ln \left( q \right)$ and $\alpha = 2.8883$ (determined from the minimization of the dissipation rate, neglecting the inertial correction). The results of the fit are presented in table 1 for each individual spectrum. For the $u'$ component of the fluctuating velocity field, Laufer \cite{Laufer} measured the spectrum at five different radial positions. However, for the other two velocity components, only three radial locations were reported. In order to have the three spectra at each of the five radial locations, an interpolation scheme was used to obtain the fitting parameters at the two missing radial positions. The interpolation scheme was based on the anisotropy of each fitting parameter in the $v'$ and $w'$ directions, respect to $u'$. The anisotropy at each radial location was calculated as the ratio of the fitting parameter in the $v'$ or $w'$ over the value obtained for $u'$. For example, we obtain $B$ at the three pipe locations for $u'u'$, $v'v'$ and $w'w'$, and then we calculate the ratio $B_{v'v'}/B_{u'u'}$. Then, an interpolating curve was fitted for this ratio (anisotropy) as a function of the radial position (individual curves for each fitting coefficient), which was then used to calculate the coefficients for the missing spectra. 
Figures \ref{fig:uSpectraCenter}, \ref{fig:uSpectraR1} and \ref{fig:uSpectraR2} show the measured (circles) and fitted (continuous lines) spectra at three radial locations. 

\section{Taylor expansion} \label{taylorExpansion}
To explain the meaning of the Taylor expansion used in the main body of the article, we expand a function $\phi\left(x \right)$ around a point $x_0$

\begin{myequation}
\begin{split}
\phi\left( x \right) = \phi\left( x_0 \right) + \phi' \left( x_0 \right)\left( x - x_0 \right) + \frac{1}{2} \phi'' \left( x_0 \right)\left( x - x_0 \right)^2+...\\
or\\
\phi\left( x \right) = \phi\left( x_1 \right) + \phi' \left( x_1 \right)\left( x - x_1 \right) + \frac{1}{2} \phi'' \left( x_1 \right)\left( x - x_1 \right)^2+...\\
or\\
\phi\left( x \right) = \phi\left( x_n \right) + \phi' \left( x_n \right)\left( x - x_n \right) + \frac{1}{2} \phi'' \left( x_n \right)\left( x - x_n \right)^2+...\\
\end{split}
\end{myequation}
Since $x_i$ can be chosen arbitrarily, we can average all these equations to obtain
\begin{myequation}
\begin{split}
\phi\left( x \right) = \frac{1}{N}  \sum _{i=0}^N  \left ( \phi\left( x_i \right) + \phi' \left( x_i \right)\left( x - x_i \right) \right.+ \\
\left. \frac{1}{2} \phi'' \left( x_i \right)\left( x - x_i \right)^2+...\right)
\end{split}
\end{myequation}
Clearly, each term of this expansion represents an averaged quantity. If we now go to the continuum limit, we arrive to
\begin{myequation}
\phi\left( x \right) =  \left\langle \phi\left( y \right) \right\rangle + 
\left\langle\phi' \left( y\right)\left( x - y\right)\right\rangle + 
\frac{1}{2}\left\langle  \phi'' \left( y \right)\left( x - y \right)^2\right\rangle+...
\end{myequation}
with  
\begin{myequation}
\begin{split}
\left\langle \phi\left( y \right)\right\rangle = \frac{1}{T}\lim_{T\rightarrow\infty} \int_{-T/2}^{T/2} \phi\left( y \right) dy\\
\left\langle \phi' \left( y\right)\left( x - y\right)\right\rangle = \frac{1}{T}\lim_{T\rightarrow\infty} \int_{-T/2}^{T/2} \phi' \left( y\right)\left( x - y\right) dy\\
\left\langle \phi'' \left( y\right)\left( x - y\right)^2\right\rangle = \frac{1}{T}\lim_{T\rightarrow\infty} \int_{-T/2}^{T/2} \phi'' \left( y\right)\left( x - y\right)^2 dy
\end{split}
\end{myequation}
The integrals of the higher order terms represent, evidently, a convolution (denoted by $*$). Thus, we arrive to 
\begin{myequation}
\phi\left( x \right) =  \left\langle \phi\left( y \right) \right\rangle + 
\phi' \left( y\right)*\left( x - y\right)+ 
\frac{1}{2}  \phi'' \left( y \right)*\left( x - y \right)^2+...
\end{myequation}
This peculiar form of the Taylor expansion allows us the express any function in terms of its expected value, plus averaged corrections.


\bibliography{bibliography0}

\end{document}